\documentclass{aa}

\usepackage{wasysym}
\usepackage{booktabs}
\usepackage{multirow}

\begin{document}

\title{Effect of stellar flares on the upper atmospheres of HD~189733b and HD~209458b}

\author{J.M.~Chadney\inst{1,2}
    \and T.T.~Koskinen\inst{3}
    \and M.~Galand\inst{2}
	\and Y.C.~Unruh\inst{2}
	\and J.~Sanz-Forcada\inst{4}
}

\institute{Department of Physics and Astronomy, University of Southampton, Southampton SO17 1BJ, UK
	\and Department of Physics, Imperial College London, Prince Consort Road, London SW7 2AZ, UK
	\and Lunar and Planetary Laboratory, University of Arizona, 1629 E. University Blvd., Tucson, AZ 85721, USA
	\and Centro de Astrobiolog\'{i}a (CSIC-INTA), ESAC Campus, P.O. Box 78, E-28691 Villanueva de la Ca\~{n}ada, Madrid, Spain
}

\abstract{Stellar flares are a frequent occurrence on young low-mass stars around which many detected exoplanets orbit. Flares are energetic, impulsive events, and their impact on exoplanetary atmospheres needs to be taken into account when interpreting transit observations. We have developed a model to describe the upper atmosphere of Extrasolar Giant Planets (EGPs) orbiting flaring stars. The model simulates thermal escape from the upper atmospheres of close-in EGPs. Ionisation by solar radiation and electron impact is included and photo-chemical and diffusive transport processes are simulated. This model is used to study the effect of stellar flares from the solar-like G star \object{HD~209458} and the young K star \object{HD~189733} on their respective planets, \object{HD~209458b} and \object{HD~189733b}. The \object{Sun} is used as a proxy for HD~209458, and \object{$\epsilon$ Eridani}, as a proxy for HD~189733. A hypothetical HD~209458b-like planet orbiting the very active M star \object{AU Microscopii} is also simulated. We find that the neutral upper atmosphere of EGPs is not significantly affected by typical flares on HD~209458 and HD~189733. Therefore, stellar flares alone would not cause large enough changes in planetary mass loss to explain the variations in HD~189733b transit depth seen in previous studies, although we show that it may be possible that an extreme stellar proton event could result in the required mass loss. Our simulations do however reveal an enhancement in electron number density in the ionosphere of these planets, the peak of which is located in the layer where stellar X-rays are absorbed. Electron densities are found to reach 2.2 to 3.5 times pre-flare levels and enhanced electron densities last from about 3 to 10 hours after the onset of the flare, depending on the composition of the ionospheric layer. The strength of the flare and the width of its spectral energy distribution affect the range of altitudes in the ionosphere that see enhancements in ionisation. A large broadband continuum component in the XUV portion of the flaring spectrum in very young flare stars, such as AU Mic, results in a broad range of altitudes affected in planets orbiting this star. Indeed, as well as the X-ray absorption layer, the layer in which EUV photons are absorbed is also strongly enhanced.}


\titlerunning{Effect of stellar flares on HD~189733b and HD~209458b}

\maketitle

\section{Introduction}
The escape of atoms and ions from Extrasolar Giant Planets (EGPs) is dependent on the composition and temperature of their upper atmospheres. The main energy input into this layer is known to be XUV (X-ray and Extreme UltraViolet, EUV)\footnote{In this paper, we consider that the X-ray waveband spans $\sim0.5$ to 10~nm, the EUV from $\sim10$ to 90~nm, and the FUV, $\sim90$ to 200~nm.} radiation from the host star \citep[e.g.,][]{Rees1989}. These high-energy emissions are linked to the star's magnetic activity. Indeed younger, more active stars possess a stronger magnetic dynamo which drives more intense, and variable, XUV emissions \citep{Ayres1997,Gudel1997,Ribas2005,Gudel2007}. In previous work, we studied the way in which the stellar spectral energy distribution affects the composition of the thermosphere and ionosphere of planets orbiting K- and M-type stars \citep{Chadney2015,Chadney2016}. In the current work, we aim to investigate the effect on planetary atmospheres of another aspect of stellar magnetic activity: flares.

Flares can have a large impact on planetary ionospheres. At Earth, as a consequence of increased EUV flux during a solar flare, Total Electron Content (TEC) in the subsolar region can increase by up to 30\% over time-scales of a few minutes \citep{Tsurutani2009}, whereas TEC enhancements by a factor of 6 have been observed in the E-region of Mars' ionosphere during a solar flare in May 2005 \citep{Haider2009}. \citet{Mendillo2006} measured enhancements in the ionosphere of Mars of up to 200~\% during a flare in April 2001, and noted that given the low altitude of the increase in electron density, there was a larger intensification of the solar X-ray flux than the EUV flux. Modelling has also been carried out on the effect of flares from April 2001 on the ionosphere of Mars by \citet{Lollo2012}. The authors found large ionospheric enhancements, especially in the lower ionsphere, where electron densities doubled. They highlighted the importance of including electron-impact ionisation for correctly modelling the impacts of the flare on the lower Martian ionosphere.

Solar flares are impulsive events, during which magnetic reconnection in the Sun's atmosphere causes a rapid release of energy \citep{Benz2008}. Photons emitted during these events are observed over the whole electromagnetic spectrum and increased photon rates from a solar flare typically last for a few hours. Low-mass stars of M and K type also have been observed to produce flares \citep[e.g.][]{Hawley1991,Cully1993,Walkowicz2011,France2016}. These are thought to be of similar origin to their solar counterparts, despite being typically two to four orders of magnitude more energetic \citep{Covino2001,Sanz-Forcada2002}. Young G stars are also capable of producing flares as intense as on M and K stars, but these events are more unusual. Indeed, K and M stars have deeper convective layers, meaning that to sustain a given activity level, a lower spin is required than in G stars. Therefore, K and M stars remain active for a much longer period than G stars. Flare events are mostly short lived in low-mass stars, lasting on the order of a few hours \citep{Walkowicz2011}. Intense flares on young, low-mass stars could have significant effects on the atmospheres of planets orbiting them, especially planets with very close orbits.

Not many modelling studies of flaring on exoplanetary atmospheres have been conducted. The lower atmosphere of an Earth-like planet orbiting the M-dwarf AD Leo was modelled by \citet{Segura2010}. The authors took into account both the enhanced photon flux from the flare, as well as simulating the effect of stellar protons. They found that in a planet with no magnetic field, stellar protons impacting the atmosphere can cause ozone depletion. \citet{Atri2017} modelled the effect of stellar protons on radiation levels at the surface of a planet to gauge the effect of particle events from active stars on a planet's biosphere. The author found that in the most extreme cases, these bursts of stellar protons that often follow flares can cause extinction events. \citet{Venot2016} studied the effect of flares from AD Leo on the lower atmosphere of exoplanets and found that the post-flare lower atmosphere is significantly affected and changed in composition.

We have chosen to model the effects of stellar flux increases due to flares on the upper atmospheres of two Hot Jupiter planets that have been extensively studied: HD~209458b and HD~189733b. Transit observations of these planets have indicated the presence of atoms and ions in their upper atmospheres, some of which are escaping \citep[e.g.][]{Vidal-Madjar2003,Vidal-Madjar2004,Linsky2010,LecavelierdesEtangs2010,LecavelierdesEtangs2012,Bourrier2013a,Ben-Jaffel2013,Vidal-Madjar2013,Ballester2015}. Modelling has confirmed that the extreme stellar radiation conditions in which these planets exist are likely to heat their upper atmospheres to the point where hydrodynamic escape sets in \citep{Yelle2004,GarciaMunoz2007,Koskinen2007,Koskinen2014}.

\citet{LecavelierdesEtangs2012} observed the Lyman $\alpha$ transit of HD~189733b in September 2011 about 8 hours after a flare on the host star HD~189733. They found a deeper transit than during observations taken in April 2010. They attributed the temporal variations in transit depth to changes in either the properties of the stellar wind or the escape rate of hydrogen from the planet or a combination of both. \citet{Bourrier2013} who developed a 3D model of the extended corona around HD~209458b and HD~189733b reached essentially the same conclusion regarding the observed variations. According to their study, the deeper transit in September 2011 could be caused either by an increase in the escape rate of hydrogen by an order of magnitude from April 2010 or an increase in the stellar wind proton density. Recently, however, \citet{Guo2016} proposed that the changes observed by \citet{LecavelierdesEtangs2012} could be explained by changes in the Spectral Energy Distribution (SED) of the star in the EUV. Specifically, if the ratio of the flux at 5 – 40~nm to the flux at 5 – 90~nm (parameter $\beta$ in their study) increased by a factor of about two between April 2010 and September 2011, the H/H$^+$ ionisation front would shift to higher altitudes and allow for a larger Lyman $\alpha$ transit depth in agreement with the observations.

In this work, we model the upper atmospheres of the two planets HD~189733b and HD~209458b during and after stellar flaring events. We have been careful to reconstruct stellar spectra that accurately reflect the emissions from the host stars HD~189733 and HD~209458, in quiescent and flaring intervals. We also include the case of a flare from a very young star, AU Mic, on a HD~209458b-like planet at different orbital distances. The latter case illustrates the effect of a very energetic flare on giant planet atmospheres. Section~\ref{sec:models} explains the different stellar and planetary models that have been used, and provides details of the simulations that we have run. Section~\ref{sec:flares} describes the construction of flaring spectra. In Section~\ref{sec:results}, the effects of flaring on the planetary thermospheres and ionospheres are presented and discussed. Finally, in Section~\ref{sec:pcle_heating} a brief calculation is presented in order to estimate the effects of a stellar proton event on mass loss from HD~189733b.

\section{Models} \label{sec:models}
\begin{figure}
\centering
\includegraphics[width=.4\textwidth]{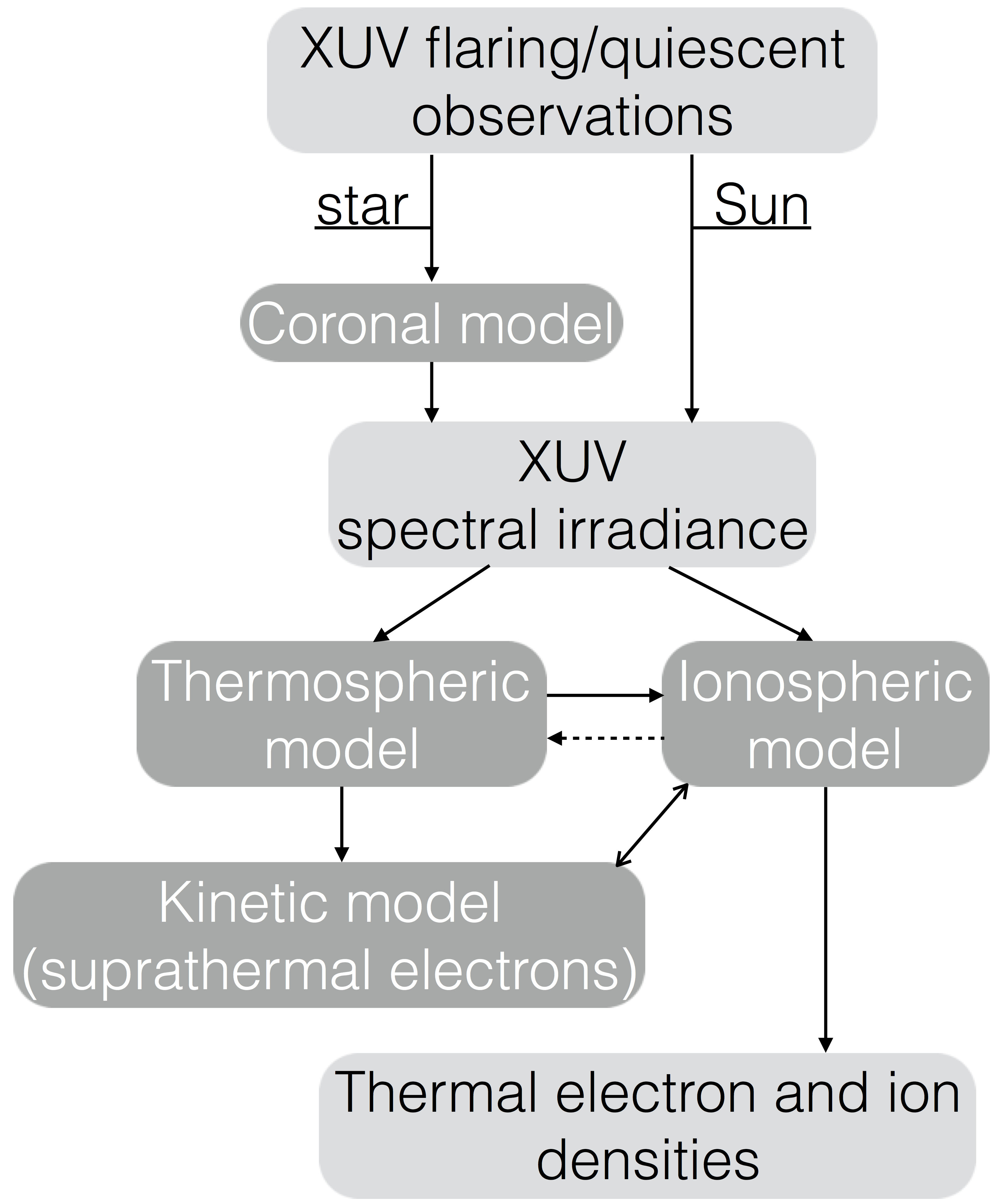}
\caption{Schematic of model elements (dark grey boxes) and inputs/outputs (light grey boxes).}
\label{fig:model}
\end{figure}

In this study, we have used a combination of models and observed stellar spectra to describe the upper atmosphere of planets HD~209456b and HD~189733b. The overall schematic of different data and model elements used to obtain ionospheric electron and ion densities is shown in Fig.~\ref{fig:model}. XUV irradiances during quiescence and flaring are constructed for each star considered, using observations of stellar emission line intensities in the soft X-ray, EUV, and FUV (Far UltraViolet). Where these observations do not span the entire spectral range of interest -- in practice, for every star but the Sun -- a coronal model is optimised to reproduce the available observed line intensities to produce a continuous spectrum from 0.1 to 91.2~nm. More details on the coronal model are provided in Sect.~\ref{sec:stellar_model}, while explanations on the construction of flaring spectra for each star are given in Sect.~\ref{sec:flare_spectra}. We use the star $\epsilon$ Eridani as a proxy for HD~189733 and the Sun as a proxy for HD~209458 (see Sect.~\ref{sec:overview_simulations} for more details).

Once the XUV spectral irradiances are obtained, these are used to drive a model of the planetary upper atmosphere. The planetary model is composed of three modules (see Sect.~\ref{sec:thermo_iono_models}): a thermospheric module describing the neutral upper atmosphere, linked to an ionospheric model that calculates densities of ions and electrons, and in addition, a kinetic module. The latter describes the energy deposition of suprathermal photo-electrons and their secondaries, allowing inclusion of ionisation by electron-impact. The thermospheric model is run with either the parameters of the planet HD~209458b or HD~189733b. A comparison of the neutral upper atmospheres of these two planets is provided in Sect.~\ref{sec:thermo_hd189b_hd209b} as well as a discussion on lower and upper boundary conditions.

\subsection{Stellar spectra} \label{sec:stellar_model}
The hydrogen and helium that make up the InterStellar Medium (ISM) are strong absorbers in the EUV. At wavelengths between about 40 and 91.2~nm (the latter corresponding to the ionisation threshold of atomic hydrogen), it is very difficult to observe emissions from any star but the Sun. The same wavelengths absorbed by the ISM are absorbed in upper planetary atmospheres and can result in ionisation. Hence we need to obtain an estimate of stellar emissions in this wavelength band. We achieve this by employing a model of the stellar atmosphere \citep{Sanz-Forcada2002,Sanz-Forcada2003,Sanz-Forcada2011}. The model builds up an Emission Measure Distribution (EMD) of the stellar corona, chromosphere, and transition region using measured emission line intensities for each star over as wide a wavelength range as possible (i.e., at wavelengths where the ISM absorption is sufficiently low to allow good S/N observations) so as to include a large range of excitation and ionisation potentials. The emission line intensities are measured for each star using spectra taken with the Chandra, XMM-Newton, and ROSAT observatories (in the X-ray); EUVE (in the EUV); and FUSE (in the FUV).

\begin{figure}
\centering
\includegraphics[width=0.49\textwidth]{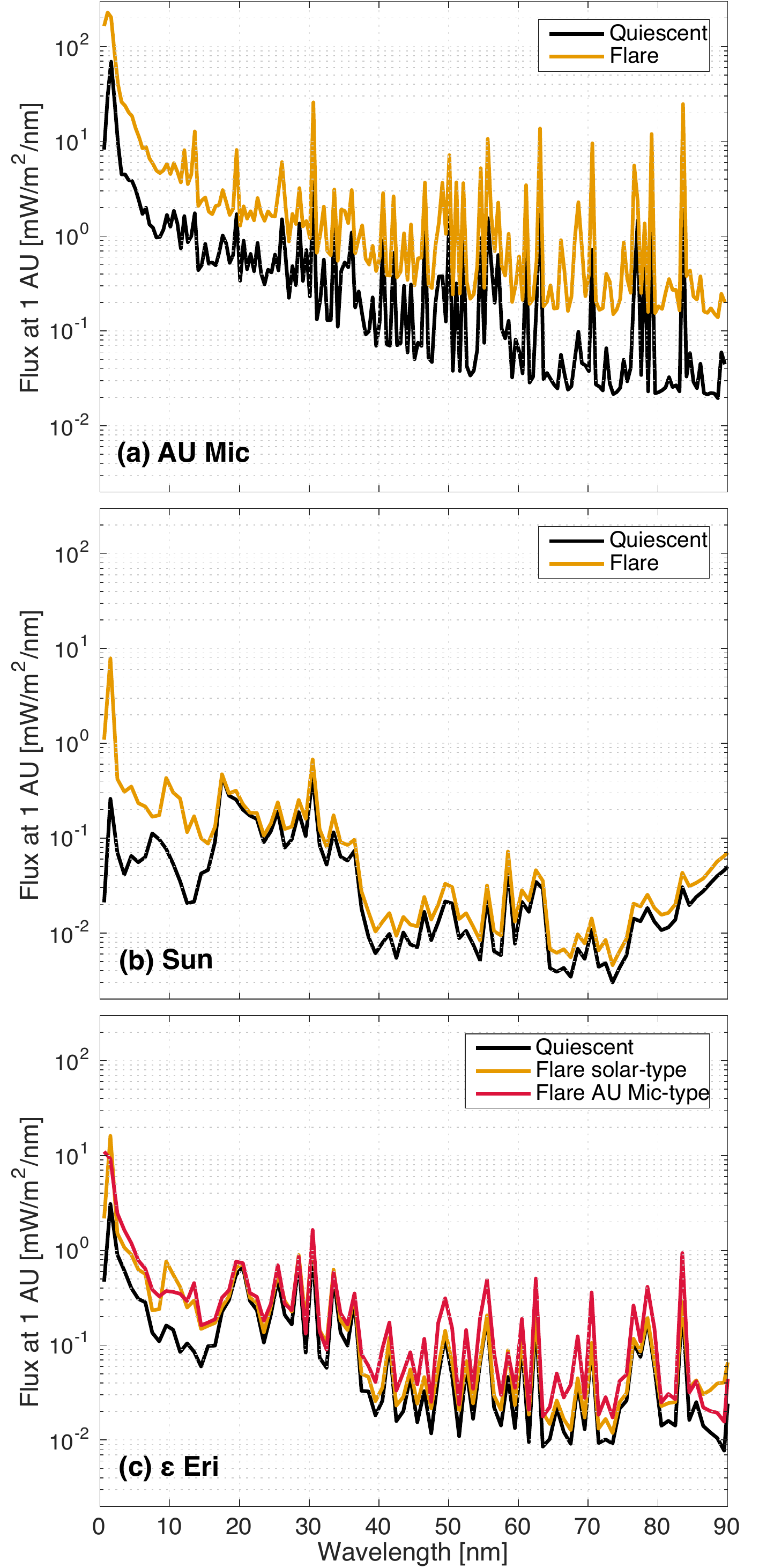}
\caption{(a) Quiescent (in black) and flaring (in orange) spectra of the star AU Microscopii, obtained from the coronal model. (b) Quiescent (in black) and flaring (in orange) spectra of the Sun (used as a proxy for HD~209458), based on measurements by the TIMED/SEE spacecraft. The flaring spectrum is the mean of four strong solar flares (see text). The quiescent spectrum is the mean of the spectra recorded a few minutes before onset of the flares. (c) Quiescent and flaring spectra of the star used as a proxy for HD~189733. The quiescent spectrum (in black) is obtained from the coronal model applied to the star $\epsilon$ Eridani. Two different flaring spectra are constructed using spectral shapes from solar flares (in orange) and the modelled AU Mic flare (in red) -- both flaring spectra are scaled to match X-ray observations of flares on the star HD~189733 (see text for further details).}
\label{fig:all_spectra}
\end{figure}

This coronal model has been used to produce spectra for stars in a quiescent state, e.g.\ for the stars $\alpha$ Cen B, $\epsilon$ Eri, AD Leo, AB Dor \citep{Sanz-Forcada2002,Sanz-Forcada2003,Sanz-Forcada2011,Chadney2016}. In the present study, we use the model-derived quiescent spectrum of $\epsilon$ Eri (as a proxy for HD~189733). Observations at different activity levels of the star AU Mic are combined with the coronal model to produce both a quiescent and, for the first time, a flaring spectrum for this star. Here we use observations of large flares on AU Mic from EUVE \citep{Cully1993} and FUSE \citep{Redfield2002}. X-ray observations of AU Mic from Chandra and XMM-Newton are not included since the flares that have been observed in these are much smaller than in the EUVE and FUSE observations (Sanz-Forcada et al., in prep.). The spectra obtained for AU Mic are shown in Fig.~\ref{fig:all_spectra}(c) and are discussed in Sect.~\ref{sec:flare_spectra}, along with quiescent and flaring spectra for HD~209458 and HD~189733, based on the Sun (see Fig.~\ref{fig:all_spectra}(b)) and $\epsilon$ Eri (see Fig.~\ref{fig:all_spectra}(c)), respectively.

\subsection{Planetary upper atmospheric model} \label{sec:thermo_iono_models}
To describe the planetary ionosphere and assess electron and ion densities, we link together three models: a thermospheric model \citep{Koskinen2013a,Koskinen2013b,Koskinen2014}, an ionospheric model \citep{Chadney2016}, and a kinetic model \citep{Moore2008a,Galand2009}. This allows the description of a one-dimensional ionosphere containing the ions H$^+$, H$_2^+$, H$_3^+$, and He$^+$, produced from the parent species H$_2$, H, and He.

The thermospheric model is driven by the stellar XUV irradiance (see Sect.~\ref{sec:overview_simulations}), using either the parameters of the planet HD~209458b or HD~189733b (given in Table~\ref{tab:HD209b_HD189b_params}) and provides neutral temperatures and number densities. The ion densities from photoionisation and subsequent chemistry are determined in the thermospheric model in order to estimate terms which depend on ionospheric conditions. This is, in particular, the case for IR cooling from H$_3^+$, which has a large effect on the neutral temperature profile. Lower and upper boundary conditions used in this model are discussed in Sect.~\ref{sec:thermo_hd189b_hd209b}.

\begin{table*}
\centering
\caption{Physical parameters of planets HD~209458b and HD~189733b and their host stars.}
\begin{tabular}{|l|ccc|cccc|}
\toprule
Star system & \multicolumn{3}{|c|}{Planet} & \multicolumn{4}{|c|}{Host star}\\ \midrule 
 & mass & radius & $a$ & mass & radius & type & age \\
 & [$M_{\text{J}}$] & [$R_{\text{J}}$] & [AU] & [$M_{\astrosun}$] & [$R_{\astrosun}$] & & [Gyr] \\
\midrule
\multicolumn{1}{|l|}{HD~209458} & 0.69$^1$ & 1.36$^1$ & 0.047$^1$ & 1.12$^1$ & 1.20$^2$ & G0 & 4.0$^3$ \\
\multicolumn{1}{|l|}{HD~189733} & 1.14$^1$ & 1.14$^1$ & 0.031$^1$ & 0.81$^1$ & 0.81$^2$ & K0 & $1-2$$^4$\\
\bottomrule
\end{tabular}
\tablefoot{Planetary masses are given in Jupiter masses ($M_{\text{J}}$) and radii are given in Jupiter radii ($R_{\text{J}}$). Stellar masses are given in terms of the solar mass ($M_{\astrosun}$) and stellar radii, in terms of the solar radius ($R_{\astrosun}$). Orbital distance is denoted $a$. The age given for HD~189733 corresponds to the `apparent' age for X-ray observations, see Section~\ref{sec:overview_simulations}. References: $^1$:~\citet{Torres2008}, $^2$:~\citet{Boyajian2014}, $^3$:~\citet{Melo2006}, $^4$:~\citet{Poppenhaeger2014}.}
\label{tab:HD209b_HD189b_params}
\end{table*}

The ionospheric model allows the inclusion of electron-impact ionisation, as well as photoionisation. It solves the coupled ion continuity equations to determine ion and electron number densities (assuming quasi-neutrality), taking into account ion photochemistry and diffusion. Electron impact ionisation is calculated using a kinetic model, which solves the Boltzmann equation. Transport, angular scattering, and energy degradation of photoelectrons and their secondaries are taken into account. Some of the model runs are for rotating planets. In these cases, the 1D coupled model is used to determine diurnal variations by calculating ion densities with altitude at a given latitude and then varying the stellar zenith angle over the course of the day. Horizontal circulation is neglected and it is assumed that vertical gradients are dominant over horizontal gradients. More details about these models and their couplings can be found in \citet{Chadney2016}.

\subsection{Overview of simulations and choice of stellar proxies}\label{sec:overview_simulations}

\begin{table*}
\centering
\caption{Details of model runs.}
\begin{tabular}{c|lcc|lcc}
\toprule
 run & \underline{star} & spectrum  & spectrum & \underline{planet} & $a$ & rotational\\
\#  & 			 & quiescent & flare & 					  & [AU]	    & period \\
\midrule
1 & HD~189733 & $\epsilon$ Eri quiet & $\epsilon$ Eri quiet + scaled Sun flare & HD~189733b & 0.031 & phase locked \\
2 & HD~189733 & $\epsilon$ Eri quiet & $\epsilon$ Eri quiet + scaled AU Mic flare & HD~189733b & 0.031 & phase locked \\
3 & HD~209458 & Sun quiet & Sun flare & HD~209458b & 0.047 & phase locked \\
4 & AU Mic & AU Mic quiet & AU Mic flare & HD~209458b & 0.2 & phase locked \\
5 & AU Mic & AU Mic quiet & AU Mic flare & HD~209458b & 1.0 & 10~hours \\
\bottomrule
\end{tabular}
\tablefoot{Numbers are used to identify each run (\#) and parameters of the star and planet modelled are provided. The flaring spectra for runs \#1 and 2 are made up of the quiet spectrum for $\epsilon$ Eri (proxy for HD~189733) onto which is added the flare excess energy of either a solar flare (\#1) or a flare on AU Mic (\#2), scaled according to X-ray observations of flares on HD~189733 (see text for more details). In the cases where the planet is phase locked with its star, the rotational period is synchronised to the planet's orbital period.}
\label{tab:run_params}
\end{table*}

We have sought to simulate three cases: the planets HD~189733b (runs \#1 and 2), HD~209458b (run \#3), and a hypothetical HD~209458b-like planet orbiting the star AU Mic (runs \#4 and 5). These runs are summarised in Table~\ref{tab:run_params}.

In the cases of the `real' planets, HD~209458b and HD~189733b, there are not enough EUV observations of their host stars to constrain the stellar coronal model in both quiescent and flaring states. Hence we have used other stars as proxies to obtain their spectra during quiet periods and during flaring. Properties of the two host stars HD~209458 and HD~189733 are given in Table~\ref{tab:HD209b_HD189b_params}. HD~209458 is a G0 main sequence star that is of similar age to the Sun, therefore we use solar XUV irradiances measured by the TIMED/SEE instrument. 

HD~189733 is a double star system comprising a K0V star, HD~189733A, and a M4V companion, HD~189733B. The hot Jupiter planet (HD~189733b) orbits the K star. There is difficulty establishing the age of this system since the two stars appear to have different activity levels. Using X-ray measurements, HD~189733A is determined to be between $1-2$~Gyr \citep{Sanz-Forcada2011,Poppenhaeger2014}, this value is in agreement with ages determined from Ca II chromospheric emission \citep{Knutson2010} and from the stellar rotation rate \citep{Poppenhaeger2014}. However, the low X-ray emission detected from the M-star companion HD~189733B indicates an age in excess of 5~Gyr \citep{Poppenhaeger2014}. It has been suggested that the higher activity of HD~189733A is linked to the hot Jupiter planet in orbit around this star \citep{Pillitteri2011,Pillitteri2014,Poppenhaeger2014}. In this scenario, the planet inhibits the spin-down of its host star, making it appear younger than it actually is when using age determinations based upon high-energy stellar emissions or rotation rate. Since it is XUV stellar emission that is absorbed in planetary upper atmospheres, for the purpose of the present study it is appropriate to consider the `apparent' age of HD~189733A, based upon X-ray observations, when looking for a proxy for this star. Therefore we have chosen the young K star $\epsilon$ Eridani to use as a proxy for HD~189733A. Note that in the rest of this paper, when referring to HD~189733, we mean the K star HD~189733A.

We have previously studied the high-energy spectrum of $\epsilon$ Eri and its effect on planetary upper atmospheres \citep{Chadney2015,Chadney2016}. However, the $\epsilon$ Eri irradiances used in these earlier publications represent a quiescent (i.e., non-flaring) period. In order to obtain a flaring spectrum for this star, we use the spectral shape of either a solar flare (run \#1) or of a flare on AU Mic (run \#2), and we scale the irradiances to match observed X-ray flares on HD~189733 \citep{LecavelierdesEtangs2012,Pillitteri2014}. The process of obtaining the flaring spectra is expanded upon in Sect.~\ref{sec:flare_spectra}.

In addition to simulating the effect of stellar flares on HD~209458b and HD~189733b, we assess the effect of the significantly more intense flares emitted by the flaring star AU Mic on a hypothetical HD~209458b-like planet orbiting at different orbital distances: run \#4 corresponds to an orbital distance of 0.2~AU and run \#5, to 1~AU. These final two runs also allow us to evaluate the effects of the planetary rotation period on the upper atmosphere. Indeed, in runs \#1 to \#4, the orbital distances are small enough that the planets can be considered to be phase-locked with their star, whereas in run \#5, the planet, located at 1~AU, is taken to have a rotational period of 10~hours, similar to that of the solar system gas giants.

\subsection{Boundary conditions and the neutral upper atmospheres of HD~209458b and HD~189733b} \label{sec:thermo_hd189b_hd209b}
\begin{figure*}
\centering
\includegraphics[width=0.98\textwidth]{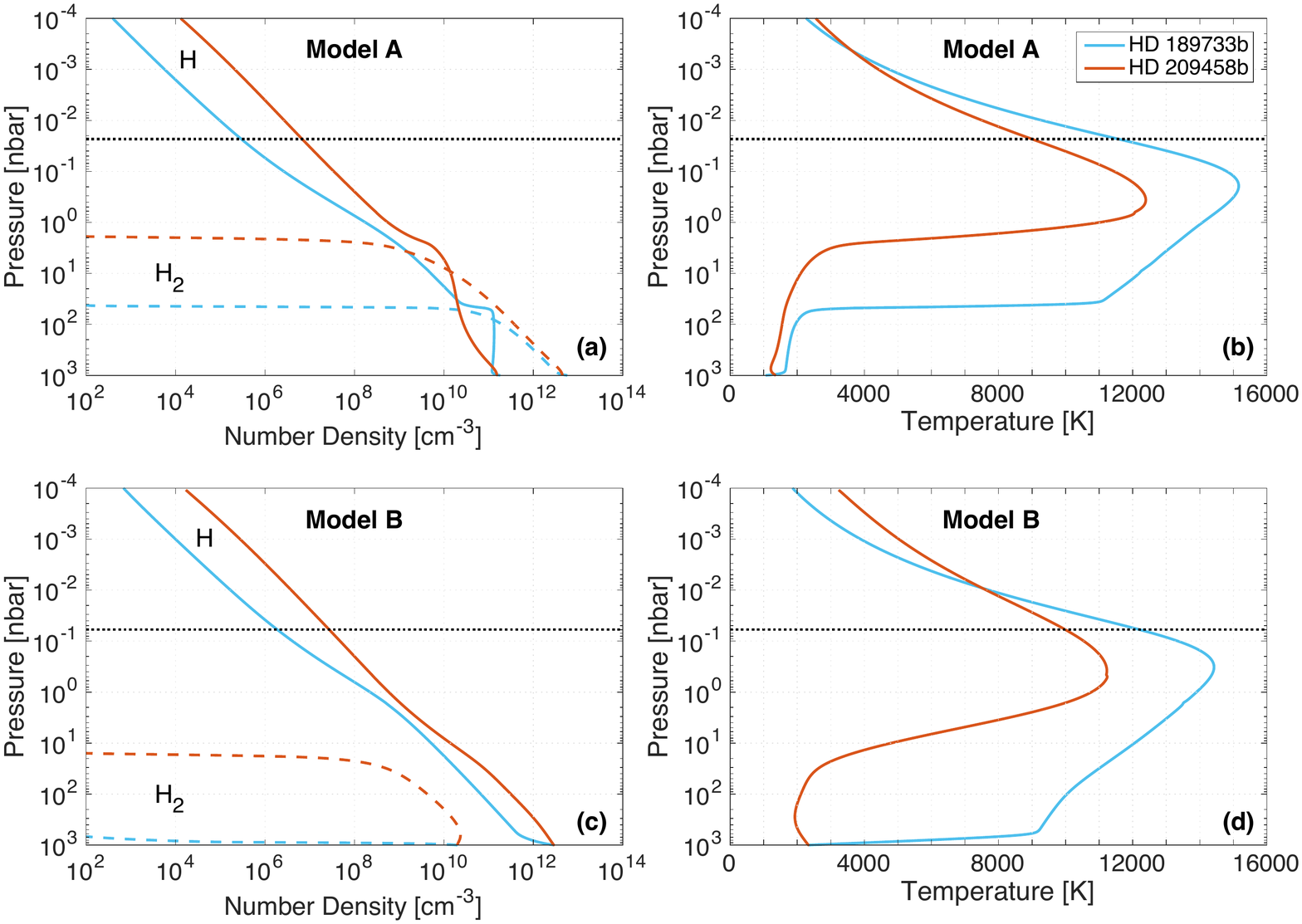}
\caption{Comparison of pressure profiles of neutral densities (left) and temperature (right) in the thermosphere of HD~209458b (red) and HD~189733b (blue) based on quiescent stellar spectra for model A with a cool lower atmosphere and model B with a hot lower atmosphere (see text). In panels a and c, the solid lines represent atomic hydrogen densities, whereas the dashed lines are molecular hydrogen. The dotted horizontal black line in each panel represents the pressure level corresponding to $2.95 R_{\text{p}}$, the lower boundary of the \citet{Bourrier2013} model.}
\label{fig:thermo_HD209b_vs_HD189b}
\end{figure*}

Before focusing on the response of the upper atmosphere to the flare in Sect.~\ref{sec:results}, we compare the neutral composition of the thermosphere of the two planets HD~209458b and HD~189733b under quiet stellar conditions. The results obtained with the quiescent stellar spectra are plotted in Fig.~\ref{fig:thermo_HD209b_vs_HD189b} for two different lower boundary conditions at the 1~$\mu$bar level that depend on the temperature profile and composition in the lower atmosphere. Many authors have assumed that HD~209458b has a stratospheric temperature inversion while HD~189733b does not \citep[e.g.,][]{Fortney2008,Showman2009}. More recent analysis of Spitzer secondary eclipse data, however, shows that a stratospheric temperature inversion is not necessary to explain observations of HD~209458b \citep{Diamond-Lowe2014}. Given the uncertainty over the presence of a temperature inversion that directly affects our lower boundary conditions, we have decided to compare lower boundary conditions without (model A) and with (model B) a stratospheric temperature inversion in the temperature-pressure (T-P) profiles that connect smoothly to the thermosphere. For models A, we adopt isothermal lower atmospheres with effective temperatures of 1353 K and 1055 K for HD~209458b and HD~189733b, respectively. For models B, we used the dayside T-P profile of HD~209458b \citep{Showman2009,Lavvas2014} for both planets, with a temperature of 2330 K at 1~$\mu$bar. In both cases, we calculate the basic equilibrium composition in the lower atmosphere and use it, together with the T-P profile, to calculate the altitude and relative abundances of H$_2$, H and He at the 1~$\mu$bar level. In this way, we set the pressure, relative abundances and temperature at the lower boundary and calculate lower boundary densities based on these values. The escape velocity is re-calculated every time step to conserve mass flux through the lower boundary. At the upper boundary, we use extrapolated upper boundary conditions that are appropriate for hydrodynamic escape with a constant slope for velocity, temperature, and densities \citep[e.g.,][]{Tian2005}. We note that our results represent globally averaged conditions valid within the Roche lobe (see Fig.~\ref{fig:thermo_HD209b_vs_HD189b}). At higher altitudes, 3D effects that shape the density distribution cannot be ignored.

In all cases, the base of the thermosphere on HD~189733b is much deeper than on HD~209458b. This is because the X-ray and high energy EUV output of $\epsilon$ Eridani (proxy used for HD~189733) is much higher than the corresponding flux of the Sun (proxy used for HD~209458) and this radiation penetrates deeper into the atmosphere. The temperature on HD~189733b is also higher than on HD~209458b, due to the higher total XUV flux of $\epsilon$ Eridani and the closer orbit of HD~189733b. For models A, H$_2$ dominates at the lower boundary where the mixing ratio of H is very low (Fig.~\ref{fig:thermo_HD209b_vs_HD189b}~a,b). The sharp thermospheric temperature gradient coincides with the level where H$_2$ dissociates, around 100~nbar on HD~189733b and around 2~nbar on HD~209458b.  Below this level, infrared cooling from H$_3^+$ and possibly other molecules keep the atmosphere relatively cool. For models B, the temperature is high enough to dissociate H$_2$ (and other molecules) almost entirely (Fig.~\ref{fig:thermo_HD209b_vs_HD189b}~c,d). As a result, the base of the thermosphere moves deeper, particularly on HD~189733b where the high-energy spectrum that penetrates to the stratosphere is stronger. In both cases, the pressure levels and thus the atmosphere are much more extended and mass loss is enhanced. For example, the mass loss rate of the hot model B for HD~209458b is more than 3 times higher than that of the cool model A. This highlights the need for future work on models of thermal structure that connect the middle atmosphere to the thermosphere. In the absence of such models, we use model A for HD~189733b and model B for HD~209458b.

\section{Stellar flares} \label{sec:flares}
This section examines stellar flares in more detail, and explains how the flaring stellar spectra are generated. For the simulations carried out in this study (see details in Table~\ref{tab:run_params}), we produce flaring (alongside quiescent) spectra for the stars HD~209458 (using the Sun as a proxy), HD~189733 (using $\epsilon$ Eri as proxy) and AU Mic.

\subsection{Flaring spectra} \label{sec:flare_spectra}
X-ray fluxes during quiescence and flaring for the Sun, HD~189733, and AU Mic are given in Table~\ref{tab:stars_fX}. For AU Mic, there are sufficient observations at different activity levels across the X-ray, EUV, and FUV wavelengths to generate a quiescent spectrum and across EUV and FUV wavelengths to generate a flaring spectrum using the coronal model (see Sect.~\ref{sec:stellar_model}) -- these spectra are plotted in Fig.~\ref{fig:all_spectra}(c). X-ray observations were not used to produce the flaring spectrum since recorded X-ray flares on AU Mic have been small compared to the flaring seen with EUVE in a July 1992 campaign \citep{Cully1993}. Therefore the flaring flux given in Table~\ref{tab:stars_fX} for AU Mic is that determined using the coronal model and corresponds to the expected X-ray flux during the July 1992 flare; note that this flare was only observed in the EUV and not in the X-ray. Flaring on AU Mic produces a large enhancement of all of the XUV emission lines. This is in contrast to solar flares, where the soft X-ray is much more enhanced than the EUV. A mean spectrum of solar flares is plotted in orange in Fig.~\ref{fig:all_spectra}(b).

\begin{table}
\centering
\caption{Stellar X-ray fluxes (0.3 -- 3~keV) at 1~AU in mW~m$^{-2}$ during quiescence ($F_{\text{X}}^{\text{qsc}}$) and during flaring ($F_{\text{X}}^{\text{flr}}$).}
\begin{tabular}{l|c|c|c}
\toprule
star & $F_{\text{X}}^{\text{qsc}}$ & $F_{\text{X}}^{\text{flr}}$ & source\\
\midrule
Sun & 0.39 & 9.3 & TIMED/SEE \\
HD~189733 & 5.7 & 21 & Swift/XRT \\
\multirow{2}{*}{AU Mic} & \multirow{2}{*}{77} & \multirow{2}{*}{360} & coronal model \\
 & & & based on EUVE data \\
\bottomrule
\end{tabular}
\tablefoot{The TIMED/SEE solar observations are detailed further in Sect.~\ref{sec:flare_spectra}; Swift/XRT fluxes for HD~189733 are from \citet{LecavelierdesEtangs2012}; the EUVE data used to constrain the flaring coronal model of AU Mic were taken in July 1992 \citep{Cully1993}. The solar values are used for HD~209458.}
\label{tab:stars_fX}
\end{table}

X-ray flares have not been measured on HD~209458. However, it is a G0-type star of similar age to the Sun, and its X-ray flux has been observed to be between that of the Sun at solar minimum and the mean Sun over the 11-year cycle \citep{Louden2017}. Therefore, for HD~209458, we use the Sun as a proxy, taking solar observations from TIMED/SEE. This instrument has been observing the Sun in the soft X-ray and EUV since 2002 and has recorded the spectra of solar flares of different intensities. Since each solar flare observed has a slightly different spectral shape, we have chosen to take the mean of four strong X-class solar flares to build the flaring spectrum used as a proxy for HD~209458. These flares are the following:
\begin{itemize}
\item X17.4 class flare (strongest recorded by TIMED/SEE) on 4 November 2003, peak at 19.53~UT,
\item X17.2 on 28 October 2003, peak at 11.10~UT,
\item X8.3 on 2 November 2003, peak at 17.25~UT,
\item X7.1 on 20 January 2005, peak at 07.01~UT.
\end{itemize}
The TIMED/SEE observations are sampled at the flare peak, which we define as the peak intensity of the He II 30.4~nm emission line. A quiescent spectrum is constructed as the mean of the TIMED/SEE observations taken a few minutes before the onset of each flare. Since the chosen solar flare events take place mostly near solar maximum, the integrated flux of the quiescent spectrum over 0.1--118~nm is 6.03~mW~m$^{-2}$ at 1~AU, higher than the mean value over the solar cycle of 4.64~mW~m$^{-2}$ \citep{Ribas2005}. These solar spectra are plotted in Fig.~\ref{fig:all_spectra}(b), where it can be seen that the most enhanced emission lines are in the soft X-ray below about 15~nm. EUV lines are also enhanced, but by a lesser amount than in the AU Mic flaring spectrum shown in Fig.~\ref{fig:all_spectra}(c).

For HD~189733, there are observations of X-ray flaring with Swift/XRT \citep{LecavelierdesEtangs2012} and XMM-Newton \citep{Pillitteri2014}. We use the Swift/XRT X-ray fluxes which were taken a few hours before a transit observation of the planet HD~189733b using HST \citep{LecavelierdesEtangs2012}. These X-ray fluxes are given in Table~\ref{tab:stars_fX}. However, the signal-to-noise ratio of observations of flares on HD~189733 is too poor to build a reliable coronal model. Therefore we use $\epsilon$ Eri as a proxy, a quiescent spectrum of which we derive using the coronal model, as detailed in \citet{Chadney2016} -- this spectrum is plotted in black in Fig.~\ref{fig:all_spectra}(a). Its integrated X-ray flux (5~mW~m$^{-2}$ in the 0.3 -- 3~keV band) matches to within measurement errors the observed X-ray flux for HD~189733, so there is no need to apply a scaling factor for the quiescent spectrum. To obtain a flaring spectrum for HD~189733, we use the spectral shape of either the solar or the AU Mic flare excess spectrum, defined as the difference between a flaring and a quiescent spectrum. This flare excess spectrum has been scaled to match the amplitude of the observed X-ray flare enhancement $F_{\text{X}}^{\text{flr}} - F_{\text{X}}^{\text{qsc}}$ observed on HD~189733 (see Table~\ref{tab:stars_fX}). 

The flaring spectra thus obtained are plotted in Fig.~\ref{fig:all_spectra}(a). The orange line in this figure is the flaring spectrum for HD~189733 calculated with the solar flare spectral shape and the red line is calculated using the flaring spectral shape from AU Mic. By construction, the integrated fluxes in the X-ray waveband are the same for the AU Mic-type flare (in red) and the solar-type flare (in orange). However, at longer wavelengths, the AU Mic-type flare shows larger enhancements of the EUV emission lines than the solar-type flare. These represent two extreme cases: the Sun is a middle age, quiet star, whereas AU Mic is a very young M dwarf of age 12~Myr \citep{Rhee2007} and as such displays intense flaring.

\subsection{Time evolution of the flare}
\citet{Walkowicz2011} studied a large number of white light stellar flares on K and M dwarf stars using the Kepler space telescope. They found that in their sample of cool stars, the peak median flaring duration is 3.5 -- 4~hours (bins of 30~minute width were used). In addition, the authors found that depending on a star's typical flare duration, the median time spent in a flaring state is between 1 and 3\%. Stars that have longer flares tend to spend less overall time flaring.

The variation in EUV flux during flares can be decomposed into phases, as described by \citet{Woods2011} using observations of the Sun. First is the impulse phase during which energy output increases rapidly, followed by a gradual phase, during which emissions generally decline slowly. Emissions from different parts of the spectrum dominate at different phases of the flare. For example, transition region emissions, like the He II 30.4\,nm line, peak during the impulse phase, whereas hot coronal lines, such as Fe XX, Fe XXIII 13.3\,nm lines peak during the gradual phase. Many of the other EUV emissions peak a few minutes after this as post-flare magnetic loops reconnect and cool.

We have taken a simplified approach to modelling the time variability of stellar flaring emissions. We consider the impulse phase to be instantaneous (i.e., the flare reaches its peak emission intensity instantaneously) and the gradual phase to be an exponential decay. We do not consider any variation with wavelength in the time profile. The stellar irradiance $I(\lambda,t)$ at each wavelength $\lambda$ and time $t$ is equal to the following:
\begin{equation} \label{eqn:irrad}
I(\lambda,t) = f(t)I_{\text{flr}}(\lambda) + (1-f(t))I_{\text{qsc}}(\lambda) \,,
\end{equation}
where $I_{\text{flr}}(\lambda)$ is the stellar spectral irradiance at the peak of the flare, $I_{\text{qsc}}(\lambda)$ is the quiescent stellar spectral irradiance, and $f$ is the normalised flare intensity, defined as:
\begin{equation} \label{eqn:flarefactor}
\begin{split}
 \text{for}\,& t<t_0,\, f(t)=0, \\
 \text{for}\,& t\geq t_0,\, f(t)=\text{exp}(-0.86\,(t-t_0)),
 \end{split}
\end{equation}
where $t$ is the time in hour and $t_0$ is the start time of the flare. This is constructed such that $f=0.05$ at $t=t_0+3.5$~hours, corresponding to the median length of cool-star flares determined by \citet{Walkowicz2011}.

Some studies have observed stellar flares that are not impulsive but instead are characterised by a slow increase in flux, followed by a decay of similar duration \citep[e.g.,][]{France2013}. We have run a model simulation using such a ``symmetric'' flare profile and obtain very similar peak ionisation enhancements to those obtained with the impulsive flare profile described above. We consider, therefore, that our conclusions are not dependent on the time profile of the flare.

\section{Effect of flaring on the upper atmosphere} \label{sec:results}

\subsection{Response of the thermosphere to flares}
Previous studies on solar system planets have shown that the ionosphere can respond to solar flares within a few minutes \citep{Tsurutani2009}, however the neutral upper atmosphere has a longer response time. We find that this response time is long enough that it is sufficient to use just the stellar quiescent spectrum within the thermospheric model, and only take into account the flare within the ionospheric model.

\begin{figure*}
\centering
\includegraphics[width=0.98\textwidth]{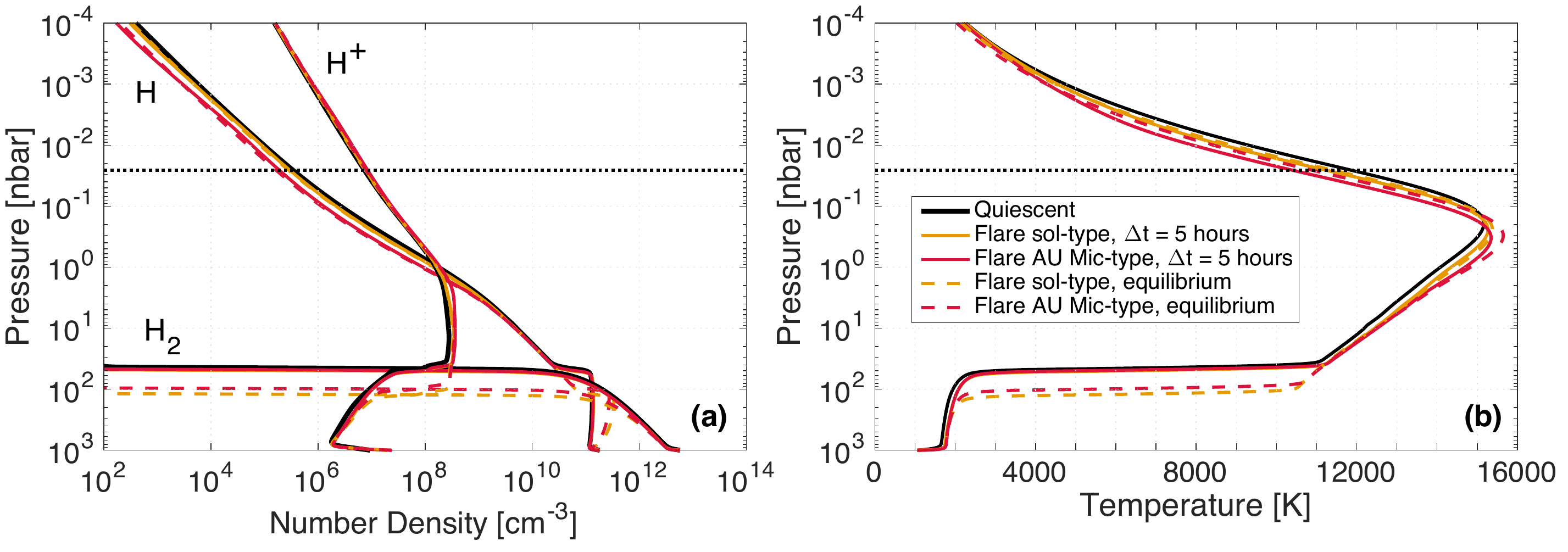}
\caption{Comparison of H$^+$, H, H$_2$ densities (a) and temperature (b) profiles in the thermosphere of HD~189733b. These are obtained from the thermospheric model driven by a quiescent spectrum (in black), by flaring spectra for a duration of 5~hours (coloured, solid lines), and by the flaring spectra to equilibrium (coloured, dashed lines). The orange and red lines correspond to the cases of solar-type and AU Mic-type flares, respectively. The dotted horizontal black line in each panel represents the pressure level corresponding to $2.95 R_{\text{p}}$, the lower boundary of the \citet{Bourrier2013} model.}
\label{fig:thermo_HD189b_qsc_vs_flr}
\end{figure*}

To test this, we ran a few different cases of the thermospheric model A on HD~189733b; the results are plotted in Fig.~\ref{fig:thermo_HD189b_qsc_vs_flr}, where panel (a) shows number densities of H, H$^+$, and H$_2$, and panel (b) shows temperature profiles. The black solid lines result from the thermospheric model driven by the quiescent spectrum of $\epsilon$ Eri, yellow lines are produced using the flaring $\epsilon$ Eri spectrum with a solar-type flare, and the red lines, with an AU Mic-type flare. For the flaring cases, the irradiance is maintained at the peak flaring value for the entire duration of the run; i.e., $f(t)=1$ in Eqn.~\ref{eqn:irrad}. Solid red and yellow lines are results from the thermospheric model after a running time of 5 hours -- longer than the flare we are simulating -- and the dashed red and yellow lines are the converged results for continuous flaring. Since the simulations in this section are run by irradiating the planets with a time-independent flux, the end-of-run flaring results shown here represent the maximum effect of the flares. This is different to the time-dependent cases studied in Sect.~\ref{sec:results_ionosphere}.

Whereas there is a considerable increase in the pressure level of H$_2$ dissociation in the case where the neutral atmosphere has been left to reach equilibrium under the continuous flare (orange and red dashed lines), this is not the case on the flaring timescale -- the solid red, yellow, and black lines of H$_2$ density are superposed in Fig.~\ref{fig:thermo_HD189b_qsc_vs_flr}(a). There is a slight difference between the different temperature profiles in Fig.~\ref{fig:thermo_HD189b_qsc_vs_flr}(b) near the temperature peak. The value of this peak remains unchanged after 5 hours, but its location has shifted lower in the thermosphere. However, given the uncertainties in the model parameters, we consider this change to be negligible and hold the neutral atmosphere fixed when we explore the impact of flaring on the ionosphere in Sect.~\ref{sec:results_ionosphere}.

The impact of the enhanced XUV flux on mass loss is of interest to the interpretation of UV transit observations. This impact is limited by the relatively short duration of the flares. On HD~209458b, the fully converged flaring model predicts a mass loss rate twice as high as the quiescent model, while the mass loss rate for the 5-hour flaring run is only up to 50\% higher than the mass loss rate for the quiescent model. For a typical middle-aged solar-type host star, we conclude that flaring activity leads to an upper limit of 1.5 for mass loss enhancement, assuming frequent flaring. On HD~189733b, the fully converged solar flare and AU Mic-type flare simulations predict mass loss rates that are, respectively, 1.8 and 2.2 times higher than the quiescent simulation. The more appropriate 5-hour runs, however, predict much weaker effects with mass loss enhancements of only 4\% and 11\% for solar and AU Mic-type flares, respectively. This is because the atmosphere does not have enough time to adjust to the flaring state before the flare has passed. We note that the lower gravity atmosphere of HD~209458b with the hot model B is more sensitive to enhanced mass loss by flaring than the relatively higher gravity atmosphere of HD~189733b with the cool model A.

\citet{Bourrier2013} proposed that the deeper Lyman $\alpha$ transit observed in September 2011 than what was seen in April 2010 can be caused either by changes in the properties of the stellar wind or an increase in the escape rate of neutral H by about an order of magnitude relative to April 2010. \citet{LecavelierdesEtangs2012} suggested that these changes could be associated with the stellar flare that they observed about 8 hours prior to the transit in September 2011. The escape rate of neutral H through a spherical surface at the radial distance of 2.95~$R_{\text{p}}$ (the lower boundary of the \citet{Bourrier2013} model, indicated by a dotted black line in Fig.~\ref{fig:thermo_HD189b_qsc_vs_flr}) increases by factor of 1.84 (from $2.4\times 10^8$ to $4.4\times 10^8$~g~s$^{-1}$) in our fully converged model of HD~189733b with the solar-type flare and by a factor 1.36 (from $2.4\times 10^8$ to $3.3\times 10^8$~g~s$^{-1}$) in the model with the AU Mic-type flare relative to the model with the quiescent spectrum. Note that the escape rates given here are for neutral hydrogen only; since the planetary outflow is mainly composed of protons, the total mass loss rates are about 3 orders of magnitude larger. The corresponding enhancements of the H escape rate in the 5-hour runs are smaller than in the fully converged runs. Therefore, we find that flaring does not change the loss rate of H by an order of magnitude. The densities of H at 2.95~$R_{\text{p}}$ in the model with solar-type flare and the model with AU Mic-type flare are $4.1 \times 10^{5}$~cm$^{-3}$ and $3 \times 10^{5}$~cm$^{-3}$, respectively, while in the model with the quiescent spectrum, the corresponding density is $3 \times 10^{5}$~cm$^{-3}$. In qualitative agreement with \citet{Guo2016}, we find that the solar-type flare, which increases the $\beta = F_{5-40\text{nm}}/F_{5-90\text{nm}}$ parameter from 0.78 to 0.81, leads to an increase in the abundance of neutral H but only by a factor of about 1.4, so a modest increase compared with what would be needed to explain the observation by \citet{LecavelierdesEtangs2012}.

\citet{LecavelierdesEtangs2012} also proposed that the observed variability can be associated with changes in the properties of the stellar wind, namely a lower proton density in April 2010.  This would lead to less efficient production of Energetic Neutral Atoms (ENAs) by charge exchange between H atoms escaping from the planet and the protons in the stellar wind \citep{Bourrier2013}. We cannot explore this idea because we do not have a 3D model of the extended corona. It would appear, however, that it should be subjected to further scrutiny because the model of \citet{Bourrier2013} does not include self-shielding of the planetary H atoms from the stellar wind by protons escaping from the planet. These protons are much more abundant above 2.95~$R_{\text{p}}$ than neutral H atoms and they present a larger collision cross section to protons in the stellar wind than neutral H \citep[cross sections given in][]{Krstic1999,Schunk2000}.

We show in Section~\ref{sec:pcle_heating} that there is a possibility that particle heating during an extreme stellar proton event could lead to an order of magnitude change in the escape rate. There is still, however, also the possibility that the observed variability is simply statistical. The signal-to-noise in the observations of HD~189733b is higher than in the observations of HD~209458b, as was pointed out by \citet{LecavelierdesEtangs2012}, but the uncertainty is still significant and the observations are affected by the absorption in the ISM and geocoronal contamination \citep{Guo2016}. To reflect this, the line-integrated Lyman $\alpha$ transit depths of $2.9 \pm1.4$~\% in April 2010 and of $5.0 \pm 1.3$~\% September 2011 are actually consistent to within 1~$\sigma$. A stronger difference between these observation dates is seen only in the blue wing of the Lyman $\alpha$ line. The false positive probability of this signal is given by \citet{Bourrier2013a} as 3.6~\%, however this was evaluated for the transit in September 2011 while it should have been evaluated for the detection of variability between the two observation dates. This problem is compounded by the fact that HD~189733 is an active star with significant spot coverage that can also affect the observed UV transit depths \citep{Koskinen2013b,Llama2016}. Repeated observations of HD~189733b are called for to confirm the detection of stellar activity-induced changes in the planetary atmosphere \citep{Fares2017}.

\subsection{Response of the ionosphere to flares} \label{sec:results_ionosphere}

\subsubsection{Planet HD 189733b irradiated by star HD 189733} \label{sec:HD189b_irrad_HD189}
We have constructed two different flaring spectra for the star HD~189733 and hence have made two different runs of the model for the planet orbiting this star (runs \# 1 and 2 in Table~\ref{tab:run_params}). Both flaring spectra have the same integrated flux in the X-ray, which matches that of the observed flare on HD~189733 \citep{LecavelierdesEtangs2012}. However in the EUV the two flaring profiles have different shapes -- the spectrum used in run \# 1 is based on the spectral distribution of a solar flare, and that used in run \#2 is based on a flare on the M dwarf AU Mic. The flare on AU Mic has a flatter energy distribution across the whole XUV range, whereas most of the energy of the solar flare is in the X-ray (see Sect.~\ref{sec:flare_spectra}). The spectrum of K dwarf $\epsilon$ Eri is used as a proxy for the quiescent state of HD~189733.

\begin{figure}
\centering
\includegraphics[width=0.49\textwidth]{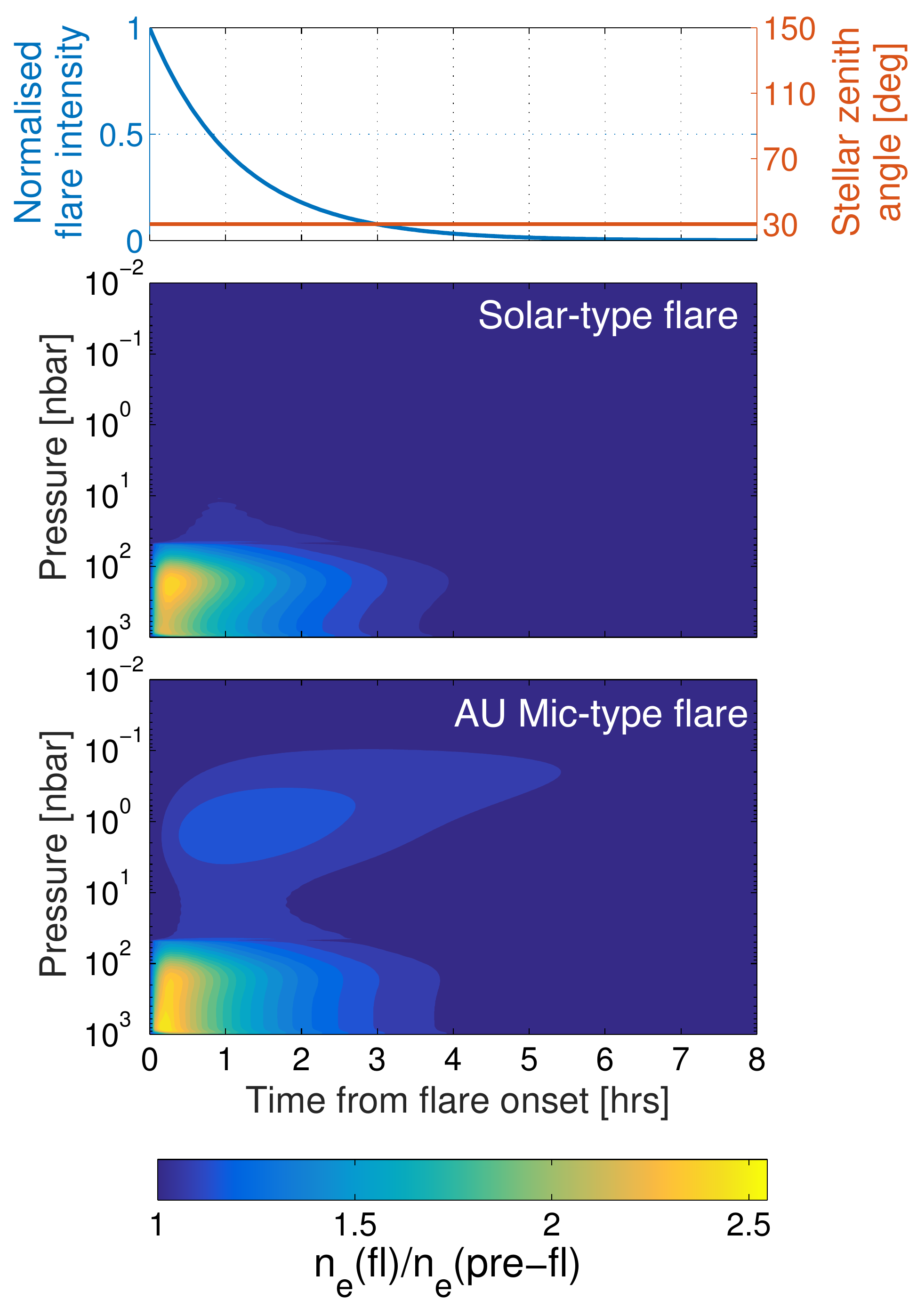}
\caption{Effects of a stellar flare on planet HD~189733b, irradiated by the star HD~189733 (using $\epsilon$ Eri as a proxy). The top panel shows the variation of the intensity of the flare (in blue) and the stellar zenith angle (in red) with time, $t=0$ representing the beginning of the flare. The middle and bottom panels show the ratio of electron density during the flare to that pre-flare, as a function of time and of pressure within the planetary atmosphere. The density ratio in the middle panel is for a planet irradiated by the solar-type flaring spectral shape and in the bottom panel, the planet irradiated by the AU-Mic flaring spectral shape. Electron densities are determined at 30$^{\circ}$ latitude and 12 Local Time (LT); the planet is considered phase-locked with its star.}
\label{fig:ne_HD189b_eeriflr_003au}
\end{figure}

The results of these flaring runs are presented in Fig.~\ref{fig:ne_HD189b_eeriflr_003au}, where the top panel shows the normalised flare intensity (in blue), defined in Eqn.~\ref{eqn:flarefactor}, and the middle and bottom panels show the enhancements in electron density during the flare, using the solar-type flare (run \# 1) and the AU Mic-type flare (run \#2), respectively. The quantity represented in the surface plots is the ratio of the electron density during the flare $n_e(\text{fl})$ to the electron density under a quiescent star $n_e(\text{pre-fl})$. The lower part of the ionosphere, at pressures higher than 50~nbar, shows the strongest response to the flare. In model A with a cool lower boundary (see Sect.~\ref{sec:thermo_hd189b_hd209b}), this region is dominated by the ion H$_3^+$, which is a short-lived ion and as such responds quickly to changes in stellar irradiance \citep{Chadney2016}. Furthermore, given the composition of the atmospheres that we consider in our model, photons of shorter wavelength are absorbed at lower altitudes. In addition, the stellar flux below 15~nm is enhanced the most by the flare. Therefore these two factors contribute to the increases in electron densities seen at the bottom of the model grid during the flare. We note that electrons are also released below the thermosphere by metals such as Mg, Na, and K, as well as through ionization of molecules such as methane, water and carbon monoxide. Modelling the effect of flares on the ionization of the lower atmosphere is beyond the scope of the present paper but it is sure to provide interesting avenues for future work, given that it can lead to changes in dynamics and energy balance \citep{Koskinen2014a}.

For run \#~1, in the high-pressure ($p>50$~nbar), H$_3^+$-dominated region, the largest increase in electron density is seen 17 minutes after the flare onset, at a pressure level of 200~nbar, where $n_e$ is multiplied by a factor of 2.5 compared to the pre-flare ionosphere. For run \#~2, the maximum electron density enhancement is by a factor of 2.5, at a higher pressure level of 800~nbar at 12~minutes after flare onset. These differences are despite the fact that the integrated X-ray flux (that is absorbed in this region of the ionosphere) for both cases \#~1 and 2 is the same. However, there are some differences in spectral shape, e.g., a stronger emission line in the spectrum of the solar-type flare at 10~nm, and more emission in the AU-Mic-type flaring spectrum between 2 and 8.5~nm. The latter are deposited deeper in the atmosphere compared with 10~nm photons, corresponding to a region with shorter timescales.

At lower pressures, where the dominant ion is H$^+$, the electron density responds far slower to the stellar flare. Since H$^+$ is a long-lived ion, the upper ionosphere also takes longer to return to its pre-flare state than atmospheric layers at pressures greater than 50~nbar. This is particularly noticeable in the bottom panel of Fig.~\ref{fig:ne_HD189b_eeriflr_003au}, which corresponds to the run with an AU Mic-type flare. The larger enhancement of EUV wavelength emission lines in this flaring scenario, compared to the solar type flare (middle panel of Fig.~\ref{fig:ne_HD189b_eeriflr_003au}), is absorbed at high altitudes. This results in a peak enhancement of electron density of 22\%, 1.25~hours after the beginning of the flare at a pressure of 1~nbar.

\subsubsection{Planet HD 209458b irradiated by star HD 209458} \label{sec:HD209b_irrad_HD209}
\begin{figure}
\centering
\includegraphics[width=0.49\textwidth]{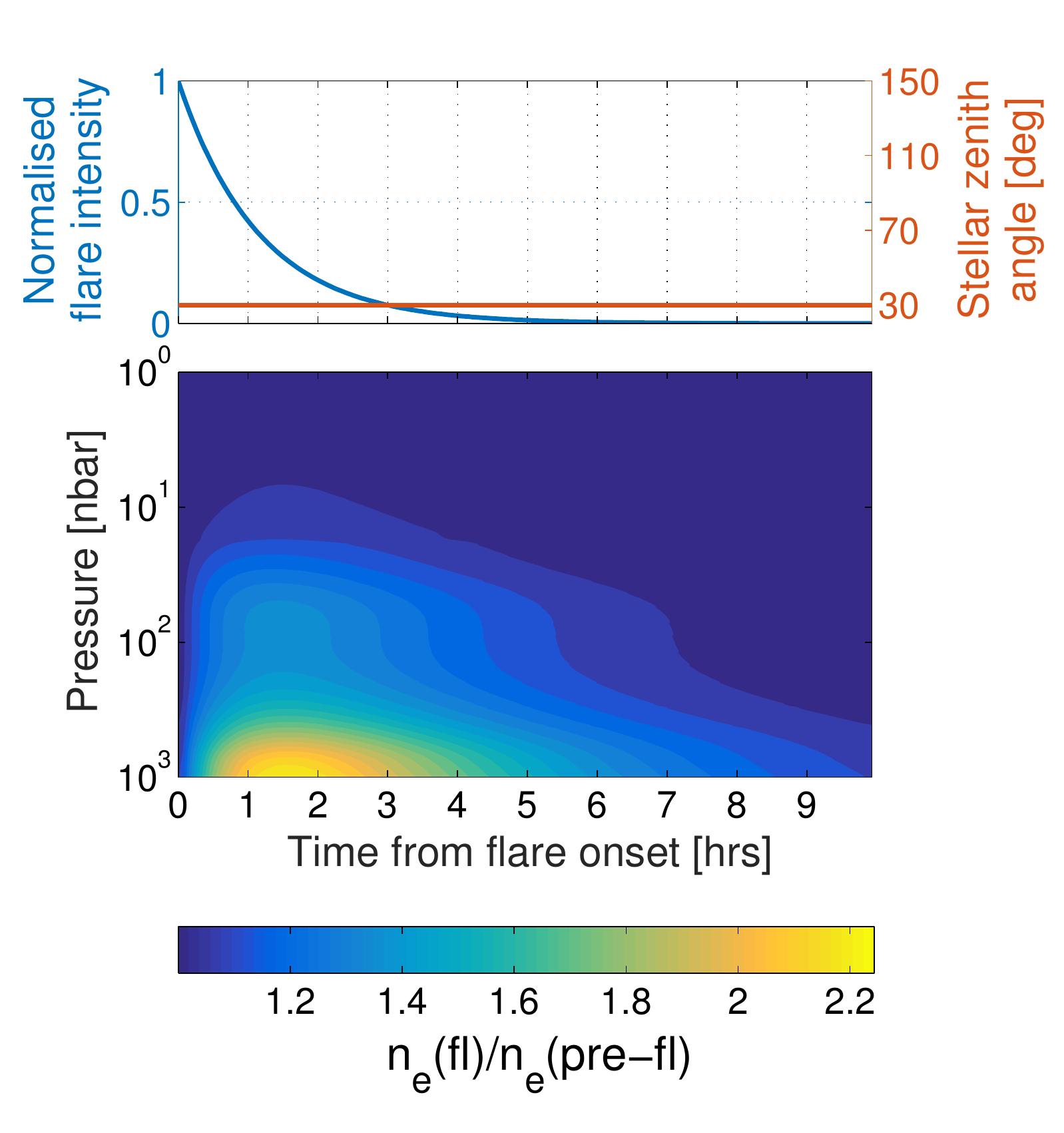}
\caption{Effects of a stellar flare on planet HD~209458b, irradiated by the star HD~209458 (using the Sun as a proxy). The top panel shows the variation of the intensity of the flare (in blue) and the stellar zenith angle (in red) with time, $t=0$ representing the beginning of the flare. The bottom panel shows the ratio of electron density during the flare to that pre-flare, as a function of time and pressure within the planetary atmosphere. Electron densities are determined at 30$^{\circ}$ latitude and 12 LT; the planet is considered phase-locked with its star.}
\label{fig:ne_HD209b_solflr_0047au}
\end{figure}

Model run \#3 is the case of the planet HD~209458b irradiated by its host star HD~209458, for which we use the solar spectrum as proxy (see Sect.~\ref{sec:flare_spectra}). The ratio of electron density during the flare to that before the flare is plotted in Fig.~\ref{fig:ne_HD209b_solflr_0047au}; note that the colour scale is different in this figure, compared to Fig.~\ref{fig:ne_HD189b_eeriflr_003au}. There are some similarities with the case of HD~189733b (see Sect.~\ref{sec:HD189b_irrad_HD189}, runs \#1 and 2): the largest ionospheric enhancement is seen at higher pressures since this is where soft X-ray photons are absorbed. Soft X-rays, at wavelengths shorter that about 15~nm, see the largest increase in flux during the flare (see Fig.~\ref{fig:all_spectra}(b)).

Nevertheless the composition of the ionosphere is different between HD~209458b and HD~189733b. In the contour plots of Fig.~\ref{fig:ne_HD189b_eeriflr_003au}, describing the planet HD~189733b, two distinct layers can be seen, with the separation pressure being about 50~nbar. This is not the case of HD~209548b (see Fig.~\ref{fig:ne_HD209b_solflr_0047au}), where only one layer is present. This is because in the latter case, using model B with a hot lower boundary (see Sect.~\ref{sec:thermo_hd189b_hd209b}), there is no H$_3^+$ layer at the bottom of the ionosphere and H$^+$ is the dominant ion throughout the modelled altitude range. H$^+$ reacts on slower timescales than H$_3^+$, hence the peak enhancement is only reached about 1.5 hours after the flare onset, as compared to 12 -- 17 minutes in HD~189733b (see Sect.~\ref{sec:HD189b_irrad_HD189}). The peak electron density enhancement at this time is $n_e(\text{fl})/n_e(\text{pre-fl})=2.2$. Another consequence of the strongest enhancement being located in a layer composed of the long-lived ion H$^+$ is that the effect of the flare lasts longer. Indeed, electron enhancements at the most affected pressure level ($p\sim1$~$\mu$bar) last up to around 9 to 10 hours after the beginning of the flare, compared to 3 to 4 hours in the case of HD~189733b.

\subsubsection{Hypothetical HD~209458b-like planet irradiated by star AU Microscopii}
For the final runs of this study, we irradiate a hypothetical planet that has the physical characteristics of HD~209458b (see Table~\ref{tab:HD209b_HD189b_params}) with the active M-dwarf star, AU Mic. We estimate the ionospheric composition for such a planet at two different orbital distances: at 0.2~AU (run \#4) and 1~AU (run \#5). At 0.2~AU, we consider the planet to be close enough to its star as to be phase-locked, whereas for the case at 1~AU, we assume a rotation period of 10~hours. Note that since these hypothetical planets orbit further away from their star than the actual HD~209458b, we consider that they do not possess a hot stratosphere \citep{Fortney2008}.

\begin{figure}
\centering
\includegraphics[width=0.49\textwidth]{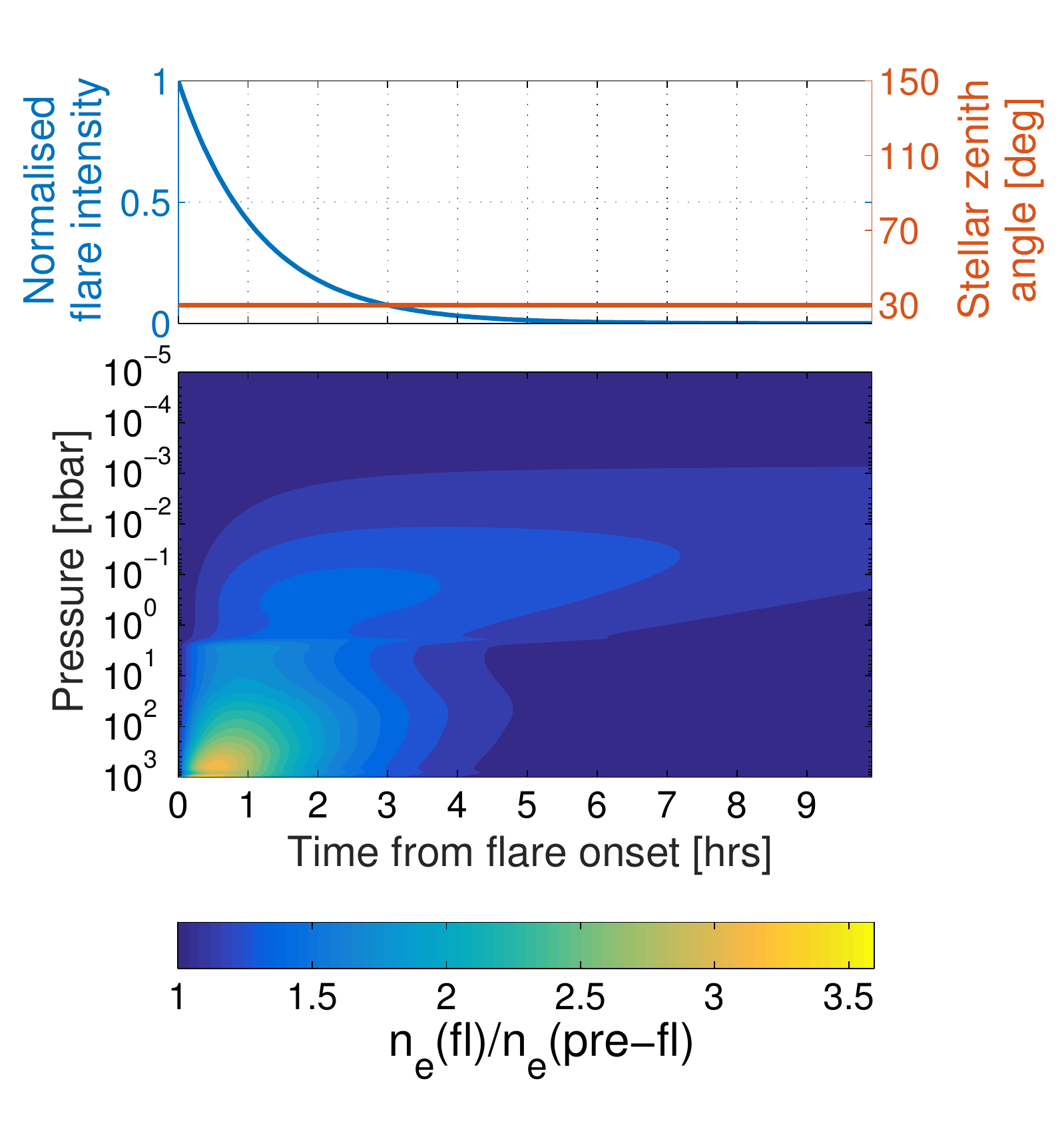}
\caption{Same as Fig.~\ref{fig:ne_HD209b_solflr_0047au} but for a stellar flare on an HD~209458b-type planet, irradiated by the star AU Mic, at an orbital distance of 0.2~AU.}
\label{fig:ne_HD209b_AUMicflr_02au}
\end{figure}

\begin{figure}
\centering
\includegraphics[width=0.49\textwidth]{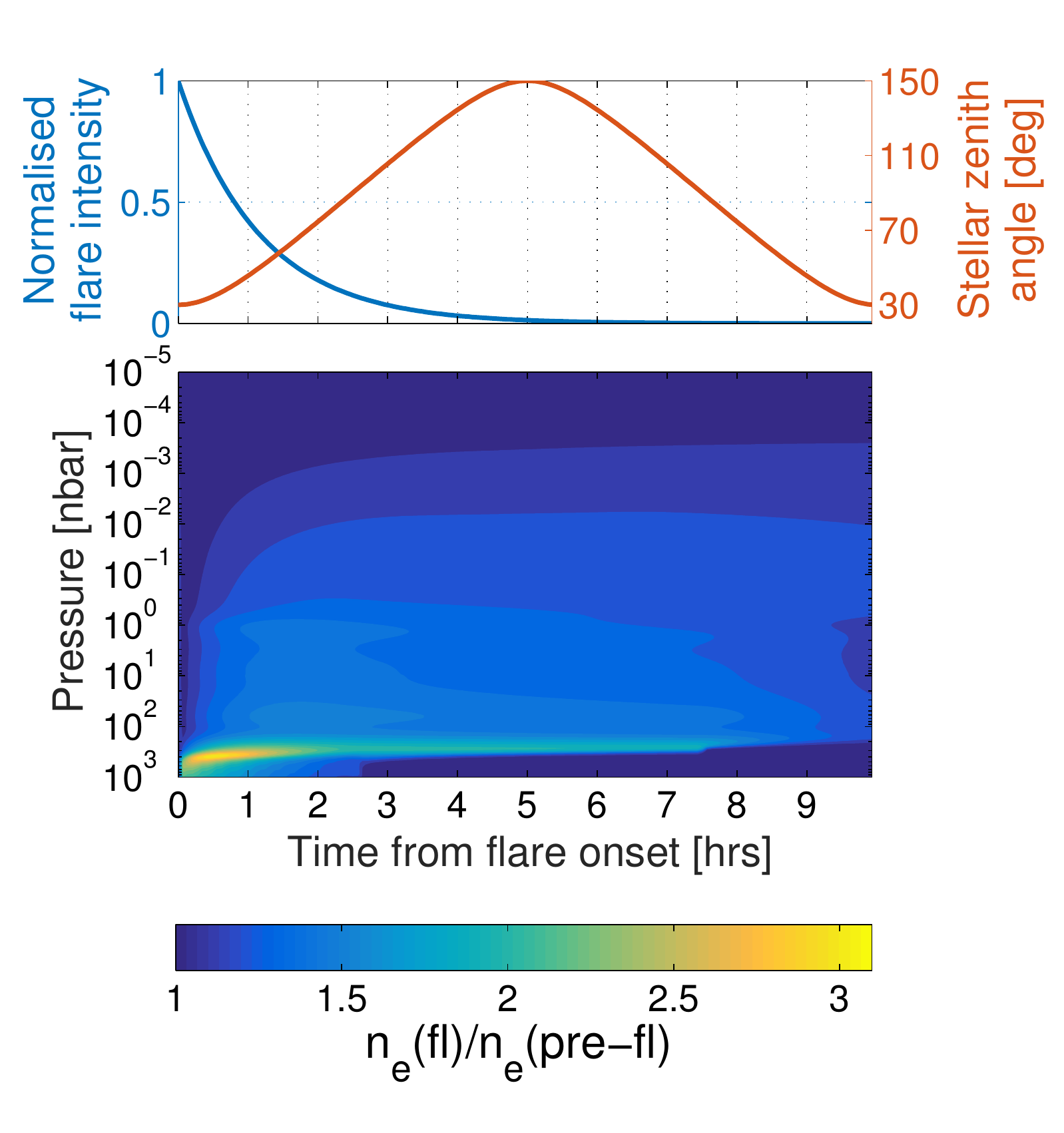}
\caption{Effects of a stellar flare on an HD~209458b-type planet, irradiated by the star AU Mic, at an orbital distance of 1~AU. The top panel shows the variation of the intensity of the flare (in blue) and the stellar zenith angle (in red) with time, $t=0$ representing the beginning of the flare. The bottom panel shows the ratio of electron density during the flare to that pre-flare, as a function of time and pressure within the planetary atmosphere. Electron densities are determined at 30$^{\circ}$ latitude and a given longitude (located at 12~LT at $t=0$), the planet is considered to rotate with a period of 10~hours. The onset of the flare corresponds to 12~LT.}
\label{fig:ne_HD209b_AUMicflr_1au}
\end{figure}

The flare studied on AU Mic is significantly more energetic than those on HD~189733 and HD~209458, detailed in Sects.~\ref{sec:HD189b_irrad_HD189} and \ref{sec:HD209b_irrad_HD209} (see Table~\ref{tab:stars_fX}). Hence we obtain greater electron density enhancements than at the other two planets, even at much larger orbital distances (note that the colour scales are different in Figs.~\ref{fig:ne_HD189b_eeriflr_003au}--\ref{fig:ne_HD209b_AUMicflr_1au}). During an intense flare on AU Mic, the stellar flux is increased over a much wider range of wavelengths than during solar flares (see Fig.~\ref{fig:all_spectra}). This results in ionospheric enhancements over a wide pressure range, as can be seen both in the planet at 0.2~AU in Fig.~\ref{fig:ne_HD209b_AUMicflr_02au} and at 1~AU in Fig.~\ref{fig:ne_HD209b_AUMicflr_1au}.

At an orbital distance of 0.2~AU from AU Mic, the ionosphere is dominated by the H$^+$ ion, due to thermal dissociation of H$_2$. There are two distinct layers of ionospheric enhancement, separated by a pressure of about 2~nbar, that react on different timescales (see bottom panel of Fig.~\ref{fig:ne_HD209b_AUMicflr_02au}). These are due to different formation and loss mechanisms of H$^+$ that operate in these altitude ranges and not to the presence of different ions. Indeed, for pressures higher than about 2~nbar, there is some H$_2$ present from which H$^+$ can be formed through photo-ionisation, electron-impact ionisation, and charge exchange reactions; for pressures below $p\sim2$~nbar, H$^+$ can only be formed from atomic H. The maximum ionospheric enhancement is at a pressure of 600~nbar, where the electron density increases by a maximum factor of 3.5 about 30 minutes after flare onset.

The picture is a little more complicated for the planet at 1~AU. This is the only case we consider of a planet that is not phase-locked with its star. Thus the effects of the flare have to be understood in conjunction with the rotating planet. We consider a planetary rotation period of 10~hours, similar to that of Jupiter and Saturn. The red curve in the top panel of Fig.~\ref{fig:ne_HD209b_AUMicflr_1au} shows the stellar zenith angle. During the model run, we follow a given longitude, at a latitude of 30$^{\circ}$. The longitude is chosen such that the beginning of the flare, at $t=0$, is at noon Local Time (LT). The resulting enhancement of electron density is shown in the bottom panel of Fig.~\ref{fig:ne_HD209b_AUMicflr_1au}. The effects of the flare on the ionosphere of this rotating planet last for a longer amount of time than in the phase-locked cases. The ionosphere of such a planet around an active star is mostly composed of the long-lived H$^+$ (see \citet{Chadney2016}), and so the additional ionisation caused by the flares lasts throughout the night over a wide range of pressures.

\section{Particle heating due to stellar proton events} \label{sec:pcle_heating}
It is possible that particle heating from a Coronal Mass Ejection (CME) that often accompanies flares could produce the changes in Lyman alpha observations seen by \citet{LecavelierdesEtangs2012} during transits of HD~189733b. While rigorously testing this hypothesis is beyond the scope of this paper, we have carried out a simplified calculation to determine whether this is likely.

Solar Proton Events (SEPs), consisting of energetic protons, can accompany fast CMEs from the Sun and heat planetary atmospheres. The highest proton fluxes from the Sun occur near solar maximum and are measured to be on the order of $10^4$~protons~cm$^{-2}$~sr$^{-1}$~s$^{-1}$ (see the NOAA SEP list at \url{ftp://ftp.swpc.noaa.gov/pub/indices/SPE.txt}). \citet{Segura2010} estimated that a large event on the young M dwarf AD Leo could produce a proton flux of $5.9\times 10^8$~protons~cm$^{-2}$~sr$^{-1}$~s$^{-1}$ for particles with energy greater than 10~MeV. To obtain an upper limit on mass loss from HD~189733b during a proton event, we use the energy-limited mass loss formula:
\begin{equation}
\dot{L} = \frac{\eta \pi r^2 F}{\Phi_0},
\end{equation}
where $\dot{L}$ is the mass loss rate, $\eta$ is the mass loss efficiency, $r$ is the radius at which the particles deposit their energy, $F$ is the particle energy flux, and $\Phi_0$ is the gravitational potential of the planet.

We assume all particles have an energy of 10~MeV and are deposited over the side of the planet facing the star (i.e., over a solid angle of $2\pi$~sr). In the presence of a strong magnetic field (and small angle between rotation and magnetic axes), solar protons are deposited throughout the polar caps, but without information on the magnetic field of HD~189733b, the size of these is difficult to estimate. In Earth's atmosphere, 10~MeV protons deposit their energy in the D-region ionosphere \citep{Turunen2009} -- below the usual main ionospheric peak. Hence, we take the radius of deposition $r$ to be equal to the bottom of the thermospheric model's altitude grid. To obtain an upper limit, we take $\eta=1$, meaning that all of the SEP energy drives mass loss. In reality this is much smaller as heating can be offset by downward conduction. In addition, SEP-induced ionisation that forms species emitting in the IR (such as H$_3^+$) could create additional cooling that may offset some heating. In all, we obtain an upper limit on mass loss from HD~189733b during a large solar-like proton event of $1.2\times 10^4$~kg~s$^{-1}$, and during a large AD Leo-type proton event of $7.0\times 10^8$~kg~s$^{-1}$.

Given that the star HD~189733 is more active than the Sun but less active than AD Leo, one might expect an upper limit mass loss rate from HD~189733b during a large stellar proton event of between $1.2\times 10^4$~kg~s$^{-1}$ and $7.0\times 10^8$~kg~s$^{-1}$. \citet{LecavelierdesEtangs2012} indicate that loss rates of $10^6$~kg~s$^{-1}$ are required for them to model the observed changes in Lyman alpha flux. It is therefore possible that a particle event is responsible for the variability seen in transit observations of HD~189733b. However it is important to note that the values derived here are upper mass loss limits of very large proton events, so for particle heating associated with stellar protons to be responsible for the \citet{LecavelierdesEtangs2012} observations, the flare seen must have been associated with an extreme proton event. Furthermore, the response of the atmosphere to a proton event is not trivial and chemical modelling is required in order to obtain reliable estimates of mass loss due to such an event.

\section{Discussion and Conclusion}
We have found that the response of the neutral atmosphere to flares from medium activity low-mass stars, such as HD~189733, and solar-like stars, such as HD~209458, is limited. Taking the median flaring time to be on the order of 3.5 to 4 hours \citep{Walkowicz2011}, the flares that we have studied on these stars are not intense enough to have a significant effect on the neutral atmosphere.

In their study of Kepler observations of flaring K and M dwarfs, \citet{Walkowicz2011} found that the median amount of time the stars were in a flaring state was only 1 to 3~\%. Much larger values were found by \citet{Hawley2014} for active M dwarfs, where as much as 30~\% of the time can be spent flaring, whereas it is between $0.01-1$~\% on inactive M dwarfs. Since HD~189733 and HD~209458 are only moderately active stars (respectively of K and G type), it is probable that they spend towards the lower end of the above range of times in a flaring state. Therefore it is unlikely that these stars would flare often enough for there to be a build-up effect of multiple successive flares resulting in a change in the neutral upper atmosphere. 

Furthermore, even if the duration of the flare is long enough for the neutral upper atmosphere to reach equilibrium, the change in neutral densities and temperature is limited on a planet like HD~189733b (see Fig.~\ref{fig:thermo_HD189b_qsc_vs_flr}). The short duration of a flare does not allow for significant enhancement in mass loss either and even in equilibrium the enhancement is at most a factor of 2. This is much less than the order of magnitude increase required by the model of \citet{Bourrier2013} to explain the deeper Lyman alpha transit seen by \citet{LecavelierdesEtangs2012} in November 2011 compared to that from April 2010. \citet{Bourrier2013} also suggested that the change in transit depth could be explained by difference in the stellar wind properties. However, this hypothesis needs to be explored further since the authors did not include self-shielding of the planetary H atoms from the stellar wind by the abundant protons escaping from the planet.

We have shown that there is a possibility that an extreme stellar proton event could be responsible for the changes in transit depth, although further modelling is required to be certain. It has also been suggested by \citet{Guo2016} that a change in the spectral energy distribution of the stellar irradiation is sufficient to explain the temporal variation in transit depth seen on HD~189733b.

\begin{figure}
\centering
\includegraphics[width=0.49\textwidth]{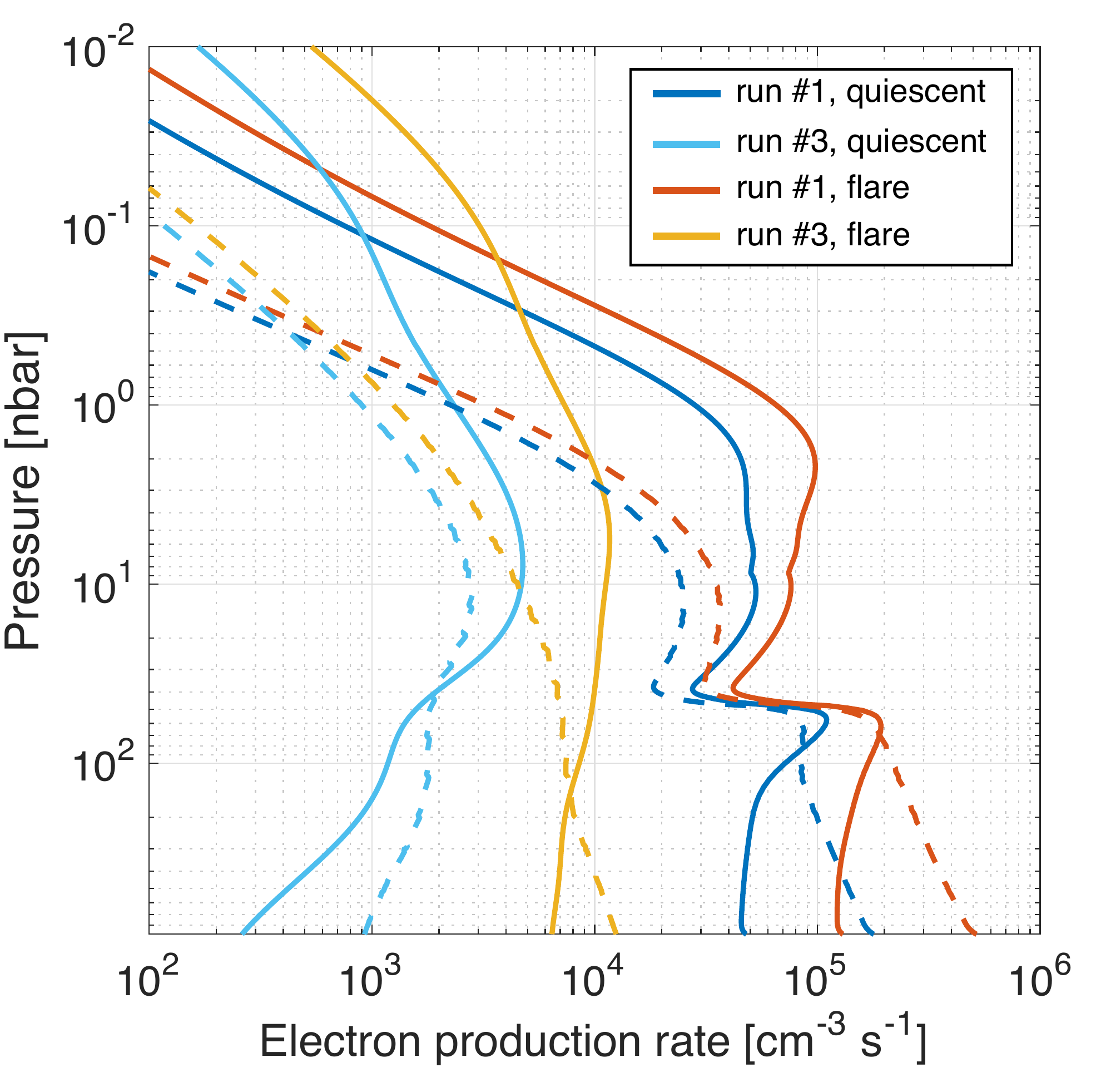}
\caption{Ionisation rates during quiescent and at the peak of the flare for runs \#1 and \#3. Solid lines represent photoionisation and dashed lines are electron-impact ionisation.}
\label{fig:ion_rates}
\end{figure}

In the various cases that we have studied, we see an enhancement of the ionosphere by a maximum factor of 2.2 to 3.5, the duration and extent in altitude of which depend on the spectral energy distribution of the flare and the composition (e.g., H$^+$ versus H$_3^+$) of the ionospheric region where enhanced flux is absorbed. The stellar soft X-ray flux increases strongly during flares. X-ray photons are absorbed in the lower part of the ionosphere, so it is this layer that sees the most enhanced ionisation during flares. This is also the region where electron-impact ionisation is dominant over photoionisation, as indicated in Fig.~\ref{fig:ion_rates}, which shows electron production rates by the two ionisation mechanisms. Electron-impact ionisation (dashed lines in Fig.~\ref{fig:ion_rates}) dominates over photoionisation (solid lines) at pressures higher than about 50 to 100~nbar.

The spectral energy distribution of the flare determines the altitude range in the planet's atmosphere that is affected by the flare. Intense flares on AU Mic display a strong continuum emission enhancement in the XUV (see Fig.~\ref{fig:all_spectra}(c)) which results in increased ionisation over a broad altitude range in the thermosphere. Large continuum enhancement during flares have also been seen in other wavebands. Strong increases in the FUV have been observed on AU Mic \citep{Robinson2001}, and also on another active M dwarf, AD Leo \citep{Hawley1991}. More recently, \citet{Kowalski2010} showed that continuum emission may be the dominant luminosity source in the Near UltraViolet (NUV) during flares. There have not been many simultaneous multi-wavelength studies of flaring stars \citep{Hawley2003,Osten2005}. There is a particular dearth of simultaneous data in the EUV which is critical for studies of exoplanetary upper atmospheres, but where measurements are difficult due to ISM absorption. More observations are needed, but if young, magnetically active stars display broadband continuum enhancements during flares, similar to AU Mic, this would mean that large altitude ranges in planetary atmospheres are affected. Programmes such as MUSCLES (Measurements of the Ultraviolet Spectral Characteristics of Low-mass Exoplanetary Systems) have begun to provide more information on the high-energy spectral shape of K and M dwarf stars, including in the EUV \citep{France2016,Youngblood2016,P.Loyd2016}. However, in these studies, the EUV flux is computed using reconstructed Lyman alpha flux assuming a spectral energy distribution similar to the Sun. This approach could lead to incorrect results for active stars. Further development of coronal models driven by multi-wavelength observations would provide the most accurate high energy spectra for active low-mass stars. Such observations should be provided by the Multiwavelength Observations of an eVaporating Exoplanet and its Star (MOVES) program which has begun to undertake a long-term study of the HD 189733 system \citep{Fares2017}.

\begin{acknowledgements}
J.M.C is funded by the United Kingdom Natural Environment Research Council (NERC) under grant NE/N004051/1. M.G. and Y.C.U.\ are partially funded by the UK Science \& Technology Facilities Council (STFC) under grants ST/K001051/1 and ST/N000692/1. J.M.C.\ has also received support from SFTC through grant ST/K001051/1 and under a postgraduate studentship, award number ST/J500616/1. J.S.F. acknowledges support from the Spanish MINECO through grant AYA2011-30147-C03-03.
\end{acknowledgements}

\bibliographystyle{aa}
\bibliography{references}

\end{document}